\documentclass[12pt]{iopart}
\expandafter\let\csname equation*\endcsname\relax
\expandafter\let\csname endequation*\endcsname\relax

\usepackage{amsmath}
\usepackage{amssymb}
\usepackage{graphicx}
\usepackage{bm}
\usepackage{url}
\def\sgn{{\rm sgn}}

\begin{document}

\title{Experimental and theoretical analysis of the upper critical field in
FSF trilayers}

\author{E~Antropov$^{1}$, Mikhail~S~Kalenkov$^{2,3}$, J~Kehrle$^{4}$, V~I~Zdravkov$^{1,4}$,
R~Morari$^{1}$, A~Socrovisciuc$^{1}$, D~Lenk$^{4}$, S~Horn$^{4}$, L~R~Tagirov$^{4,5}$, 
Andrei~D~Zaikin$^{2,3}$, A~S~Sidorenko$^{1,2}$, Horst~Hahn$^{2}$ and~R~Tidecks $^{4}$}

\address{$^{1}$Institute of Electronic Engineering and Nanotechnologies ``D. Ghi\c{t}u''
ASM, Kishinev, MD2028, Moldova}
\address{$^{2}$Institute of Nanotechnology, Karlsruhe Institute of Technology (KIT), 76021
Karlsruhe, Germany}
\address{$^{3}$I.E. Tamm Department of Theoretical Physics, P.N. Lebedev Physical
Institute, 119991 Moscow, Russia}
\address{$^{4}$University of Augsburg, D-86135 Augsburg, Germany}
\address{$^{5}$Solid State Physics Department, Kazan Federal University, Kazan, 420008,
Russia}
\ead{vladimir.zdravkov@physik.uni-augsburg.de}
\begin{abstract}
The upper critical magnetic field $H_{c2}$ in thin-film FSF trilayer spin-valve cores  
is studied experimentally and theoretically in geometries perpendicular and parallel
to the heterostructure surface. The series of samples with variable thicknesses 
$d_{\mathrm{F1}}$ of the bottom  and $d_{\mathrm{F2}}$ of the top  
Cu$_{41}$Ni$_{59}$ F-layers are prepared in a single run, utilizing a
wedge deposition technique. The critical field $H_{c2}$ is measured in the
temperature range $0.4-8$~K and for magnetic fields up to 9 Tesla.
A transition from oscillatory to reentrant behavior of the superconducting 
transition temperature versus 
F-layers thickness, induced by an external magnetic field, 
has been observed for the first time. In order to properly interpret 
the experimental data, we develop a quasiclassical theory, enabling one
to evaluate the temperature dependence of the critical field 
and the superconducting transition temperature for an arbitrary
set of the system parameters. A fairly good agreement between our experimental data
and theoretical predictions is demonstrated for all samples, using a single set 
of fit parameters. This confirms adequacy of the Fulde-Ferrell-Larkin-Ovchinnikov 
(FFLO) physics in determining the unusual superconducting properties of 
the studied Cu$_{41}$Ni$_{59}$/Nb/Cu$_{41}$Ni$_{59}$ spin-valve core trilayers.
\end{abstract}

%Uncomment for PACS numbers title message
%\pacs{74.25.Dw, 74.45.+c, 75.75.Cd}
% Keywords required only for MST, PB, PMB, PM, JOA, JOB? 
%\vspace{2pc}
%\noindent{\it Keywords}: Article preparation, IOP journals
% Uncomment for Submitted to journal title message
%\submitto{\JPA}
% Comment out if separate title page not required
\maketitle 

\section{Introduction}

The upper critical magnetic field $H_{c2}$ of an isotropic type-II
superconductor generally obeys a linear temperature dependence in the vicinity
of the superconducting transition temperature $T_{c}$ \cite{Werthamer66}.
Deviations from the linear $T$-dependence of $H_{c2}(T)$ are usually
ascribed to inhomogeneities distributed in the sample, which can broaden
the resistive transitions $R(T)$ and $R(H)$ \cite{Zwicknagl84,Larkin71}.
The temperature dependence of $H_{c2}$ is also known to be sensitive to the
orientation of the magnetic field, if the superconducting nucleus size
becomes comparable to the characteristic dimensions of the structure (see, for
example, Refs. \cite{Lee97,Drechsler00}). In particular, artificially
prepared metallic multilayers (ML) consisting of alternating superconducting
(S) and normal metal (N), or of S and insulating (I) layers, or even of two
different superconductors, S and S$^{\prime }$, show nonlinear $H_{c2}(T)$
dependences (see, for example, an early review \cite{Jin89}).

Among a variety of layered superconducting systems,
superconductor-ferromagnet (S/F) metallic hybrids attract special attention
because of rich physics of these objects \cite%
{Buzdin05,Efetov05,Pokrovskii05} as well as promising perspectives for
their applications in superconducting spintronics \cite%
{Kupriyanov04,Ryazanov1,Ryazanov2}. Based on the unique properties of S/F
hybrids, several kinds of device physics were proposed such as
proximity-effect superconducting spin-switching \cite%
{Beasley97,Tagirov99,Buzdin99,Gu02,Fominov10} and Josephson current
switching \cite{Karmin1,Houzet,Karmin2} actuated by an external magnetic
field. Very recently, magnetic-field-controlled superconductivity switching
in the S/F proximity systems \cite{Luo,Leksin1,Leksin2,Leksin3,Zdravkov12} and
the Josephson current switching in the S/F/S junctions \cite{Ryazanov12,Ryazanov12-2}
were demonstrated experimentally.

The influence of the magnetic field on superconductivity in S/F hybrid structures
was pointed out in theoretical studies\cite{Radovic88,Kuboya}. It was demonstrated 
that in order to evaluate correctly their
critical magnetic fields it is necessary to account for the magnetic field
penetration into both, the superconducting and ferromagnetic layers. An
approximate single-mode analysis for the critical field \cite{Radovic88} was
further developed in Ref. \cite{Yoksan06}, where rigorous solution of
the quasiclassical Usadel equations \cite{Usadel70} was worked out both for
SF-bilayers and for SF-multilayers. These theoretical studies of the critical fields are
important not only because of possible influence of the pivoting field on
the onset of superconductivity in the system, but in view of justifying 
the consistency of the FFLO \cite{FF,LO} physics, implemented into 
the Usadel formalism, with the actual experiments. 
Moreover, more experimental data collected here reduce ambiguities in 
fitting procedures for large number 
of physical parameters (from 6 to 12, see below).

Measurements of the upper critical fields 
in S/F multilayers since the very beginning of their studies served as 
an important proof of coupling between the layers \cite{Chien}. Samples with decoupled S-layers
are described by the theoretical approach of Ref. \cite{Radovic88}. In a magnetic field parallel to the layers a two dimensional (2D) behaviour is observed close to $T_{c}$ \cite{Chien,Verbanck,Aarts, Mattson97}. In the case of a coupling, a 2D-3D crossover occurs \cite{Chien,Verbanck2}. There are also  studies of $H_{c2}$ in ferromagnetic 
alloy - superconductor heterostructures \cite{Armenio,Huang,Cirillo}, where 
2D-3D crossover, the flux pinning mechanism and the anisotropy coefficient, 
$\gamma_{GL}$=$H_{c2}^{\parallel }(0)$/$H_{c2}^{\perp }(0)$, were deduced. 
In the case of S-layers coupling, the difference in the phase of the superconductig order parameter between S-layers can vary what
 significantly affects on $T_{c}$ and complicates the experimental detection 
of oscillatory $T_{c}(d_{F})$ behavior \cite{Prishepa}. 
Therefore, to explore experimentally the evolution of the $T_{c}(d_{F})$ dependence
 from oscillatory type to reentrant one under applied magnetic field 
we have chosen the system with sole superconducting layer. 
Our experiments, for the first time, refer to the regime of reentrant 
superconductivity to which we could drive our samples from the oscillatory 
regime, by applying an external magnetic field (see below).                                                          

Being interested in superconducting spin-valve physics, proposed by the
Beasley group \cite{Beasley97} and later developed in Refs.
\cite{Tagirov99,Buzdin99,Gu02,Fominov10}, here we present the results 
of our measurements of a critical magnetic field 
$H_{c2}$ in Cu$_{41}$Ni$_{59}$/Nb/Cu$_{41}$Ni$_{59}$ spin-valve cores 
for the parallel and perpendicular to the sample plane geometry. 
Since our samples exhibit expressed oscillatory $T_{c}(d_{F})$ behavior 
as a function of the CuNi alloy thickness at zero magnetic field, 
for our $H_{c2}(T)$ measurements we have chosen several characteristic 
points at the steeply descending, minimum, increasing and asymptotic 
segments of this $T_c(d_F)$ dependence. In addition, in the $T_c(d_F)$ 
dependence, the transition from oscillatory to reentrant behavior, 
driven by the external magnetic field, has been observed for the first time. 
A detailed analysis of the parallel and perpendicular critical fields 
as well as superconducting transition temperatures is given within 
the framework of the quasiclassical Usadel equations formalism, which 
appears to be most suitable one for the short mean-free-path materials 
utilized in our samples.

Our paper is organized as follows. In Section II we present details of 
fabrication and characterization procedures of our FSF trilayer 
samples and describe measurements
of their temperature dependent critical magnetic field $H_{c2}(T)$. Section
III is devoted to theoretical analysis of the problem within the framework
of quasiclassical Usadel equations. Section IV presents a detailed comparison between
our experimental data and theoretical calculations as well as a brief discussion
of our key observations.

\section{Experiment}

\subsection{Thin film deposition and sample preparation}

The three-layer Cu$_{41}$Ni$_{59}$/Nb/Cu$_{41}$Ni$_{59}$ structures were
grown on commercial silicon substrate
utilizing a magnetron sputtering machine by the Leybold Company, model Z-400.
Three mounted targets allowed fabrication of
the whole multi-structured samples within one cycle of sputtering without
breaking vacuum in the chamber. The interfaces between the constituent thin
films were found extremely clean \cite{Sidorenko10}. Since our study
requires a set of specimens with identical thickness of the superconducting
layer of Nb and varied thickness of the ferromagnetic layers, the method of
growing wedge-shaped films was applied. It allowed us to obtain a set of up
to 40 specimens prepared in a single run on a long substrate under identical
vacuum conditions.

Actually, a commercial silicon wafer was cut into pieces of  size $%
7\times 80$ mm$^{2}$, cleaned and put into the vacuum chamber. After that
the chamber was evacuated with a turbo-molecular pumping system to a base
pressure of $3\times 10^{-6}$ mbar, and the sputtering was done in the atmosphere of
argon ($99.999\%$ purity) with the pressure of $8\times 10^{-3}$ mbar. At
the beginning, each target ($75$ mm in diameter) was presputtered during about $10$
minutes in order to clean its surface from possible oxides and absorbed
gases.

The first Cu$_{41}$Ni$_{59}$ layer was grown on an amorphous silicon buffer
film deposited just before, to isolate the structure from gases absorbed by a
natural oxide on a surface of the substrate. To provide superior homogeneity
of the Si and Nb film thickness over the size of the substrate the
"spray" deposition technique was employed \cite{Zdravkov06,Sidorenko09}. To
realize this technique a magnetron was moved along the substrate by a custom
motorized setup providing a uniform coverage of the substrate surface with the
sputtered material.

\begin{figure}[tbp]
\centerline{
\includegraphics[width=80mm]{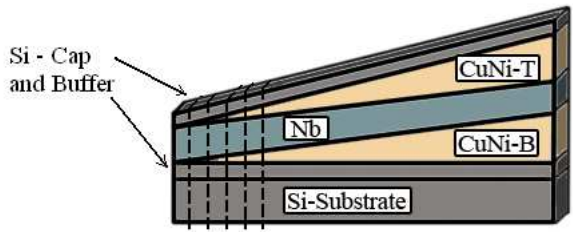}
}
\caption{(Color online) The sketch of our trilayer FSF sample with top (CuNi-T), 
bottom (CuNi-B) CuNi ferromagnetic layers and superconducting (S) Nb layer.}
\label{trilayer-fig}
\end{figure}

To deposit a wedge-shaped Cu$_{41}$Ni$_{59}$ layer, the target was
positioned with the symmetry axis just above the substrate edge. A wedge
shape of the ferromagnetic alloy was obtained due to the native gradient of
sputtering aside of the symmetry axis. The deposition rate on the thick end
of the wedge was about 3-4 nm/sec. As in the earlier experiments 
\cite{Sidorenko10,Zdravkov11}, we operated the magnetron in the ac regime to keep
the concentration of nickel about 59\% in the deposited CuNi alloy films 
(hereafter the Cu$_{41}$Ni$_{59}$ alloy will be referred as CuNi alloy). As we
already pointed out, the superconducting niobium layer was grown with the Nb
target moving along the substrate. In this way the reduced effective
deposition rate of about 1.3 nm/sec was achieved \cite{Sidorenko10,Kehrle12}. 
Subsequently, the second wedge-shaped ferromagnetic layer was sputtered on
top of the Nb film with the thickness of the both CuNi layers
varied from 2 to 50~nm (more accurate data are presented in Sec.~II). In
order to protect our FSF structure from oxidation in ambient
conditions it was capped by a thin silicon layer. The sketch of the whole FSF 
structure is presented in Fig. \ref{trilayer-fig}.

To obtain a series of FSF strips with varying CuNi
layer thicknesses $d_{F}$ to be used in our $H_{c2}(T,d_{F})$ measurements,
samples of equal width (about 2.5 mm) were sequentally cut perpendicularly to the wedge
gradient and correspondingly numbered resulting in the FSF3 batch. 
Then, aluminum wires of 50~$\mu$m in diameter were attached to
the strips by ultrasonic bonding for four-probe resistance measurements.

\subsection{Thickness and composition characterization}

The thickness of the each layer in our FSF structure was determined
by means of Rutherford Backscattering Spectrometry (RBS). He$^{++}$ ions
were accelerated to energies of 3.5 MeV by a tandem accelerator. In order to
avoid channeling effects the samples were tilted azimuthally by 7$^{\circ }$, 
and the backscattered ions were detected at the angle of 170$^{\circ }$
with respect to the incident beam.

\begin{figure}[tbp]
\centerline{
\includegraphics[width=80mm]{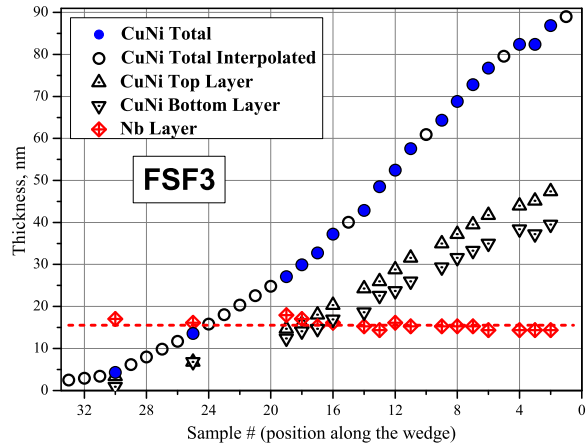}
}
\caption{(Color online)  Thicknesses of the layers of FSF3 batch evaluated by 
RBS spectrum over the whole strip-like specimens.
}
\label{FSF3-RBS-all}
\end{figure}

The RBS spectra fitting procedure is described in Ref. \cite{Kehrle12}. 
The results of the RBS evaluation are presented in Fig.~\ref{FSF3-RBS-all}.
Depending on the sample, the graph shows almost linearly increasing thicknesses of
both copper-nickel layers from about 1 to 48~nm. The niobium layer
thickness is nearly constant of around 15.5~nm (dashed line in Fig.~\ref{FSF3-RBS-all}) 
with slight tendency to increase for largest sample numbers. The thicknesses of both 
CuNi alloy layers were found to be nearly the same only for intermediate range 
of the samples (approximately for Nr. 20$\div{27}$). 
However, the deviation of the CuNi layers thicknesses over the mean value 
mainly did not exceed 10 \%.  

\subsection{Critical fields measurements}

\begin{figure}[tbp]
\centerline{
\includegraphics[width=80mm]{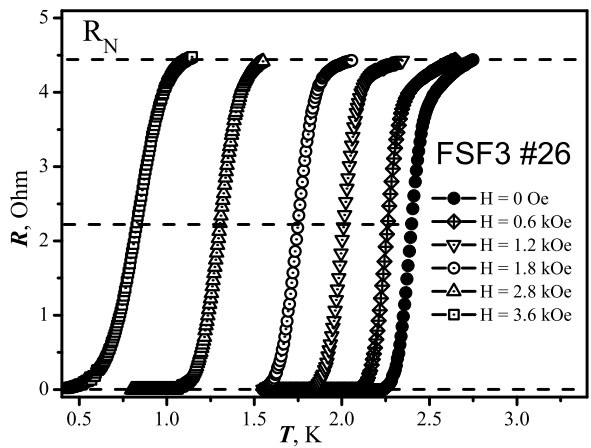}
}
\caption{(Color online) Resistive transitions for the sample \#26 of the batch FSF3 
in the presence of the magnetic field.}
\label{RT-FSF3-26}
\end{figure}

The critical field measurements were performed in a $^{4}$He cryostat
equipped with a superconducting solenoid providing magnetic fields up to $17$
\thinspace T. The temperature was controlled in the range of $0.4-8$~K with
an accuracy of about $1$~mK. Resistivity measurements were performed in 
Oxford Instruments $^{4}$He-$^{3}$He Heliox insert, using the conventional 
four-probe method. In order to avoid thermoelectric voltages, alternating 
of the current polarity was performed during our measurements. 
The resistive transitions for the sample \#26 are displayed in 
Fig. \ref{RT-FSF3-26} for different values of the magnetic field. 
The critical temperatures $T_{c}$ assigned to the applied field, 
\textit{i.e.} the upper critical magnetic field $H_{c2}$, were determined as
midpoints of the resistive transitions $R(T)$. The estimated accuracy of the $T_c$ 
evaluation is within a few mK.
The critical values of the magnetic field were measured
for two field orientations -- perpendicular and parallel to the sample
plane, $H_{c2}^{\perp }(T)$ and $H_{c2}^{\parallel }(T),$ respectively. Both
the critical temperature values and the slope of the curves $H_{c2}^{\perp
}(T)$ in the vicinity of $T_{c}$ are listed in Table \ref{sample-tbl}.

\begin{table}
%[tbp]
\caption{Sample characteristics together with the results of our
measurements. Total thickness of the ferromagnetic layers is denoted as $d_{
\mathrm{CuNi}}$. The critical temperature, ${T_c}$, is given for \textit{H} = 0. }
%\begin{ruledtabular}
%\begin{tabular}{llllllllll}
%\begin{indented}
%\item[]
{\small
\begin{tabular}{@{}llllllllll} 
\br
Sample number& 33 & 31 & 28 & 26 & 20 & 15 &  10 & 5 & 1\cr
\mr
$d_{\mathrm{CuNi}}$
           & 2.5 & 3.4 & 6.6 & 10.8 & 23.9 & 39.6 & 60.1 & 79.5 & 89.0\cr
$T_c$ (K)  & 6.73 & 6.7 & 5.25 & 2.40 & 2.69 & 2.86 & 2.98 & 3.00 & 3.35\cr
$\partial H_{c2}/\partial T$ (T/K)&-0.522&-0.469&-0.477&-0.236&-0.216&-0.221& 
            -0.217&-0.217&-0.216\cr
\br
\end{tabular}}
%\end{indented}
%\end{ruledtabular}
\label{sample-tbl}
\end{table}

\begin{figure*}[tbp]
\includegraphics[width=80mm]{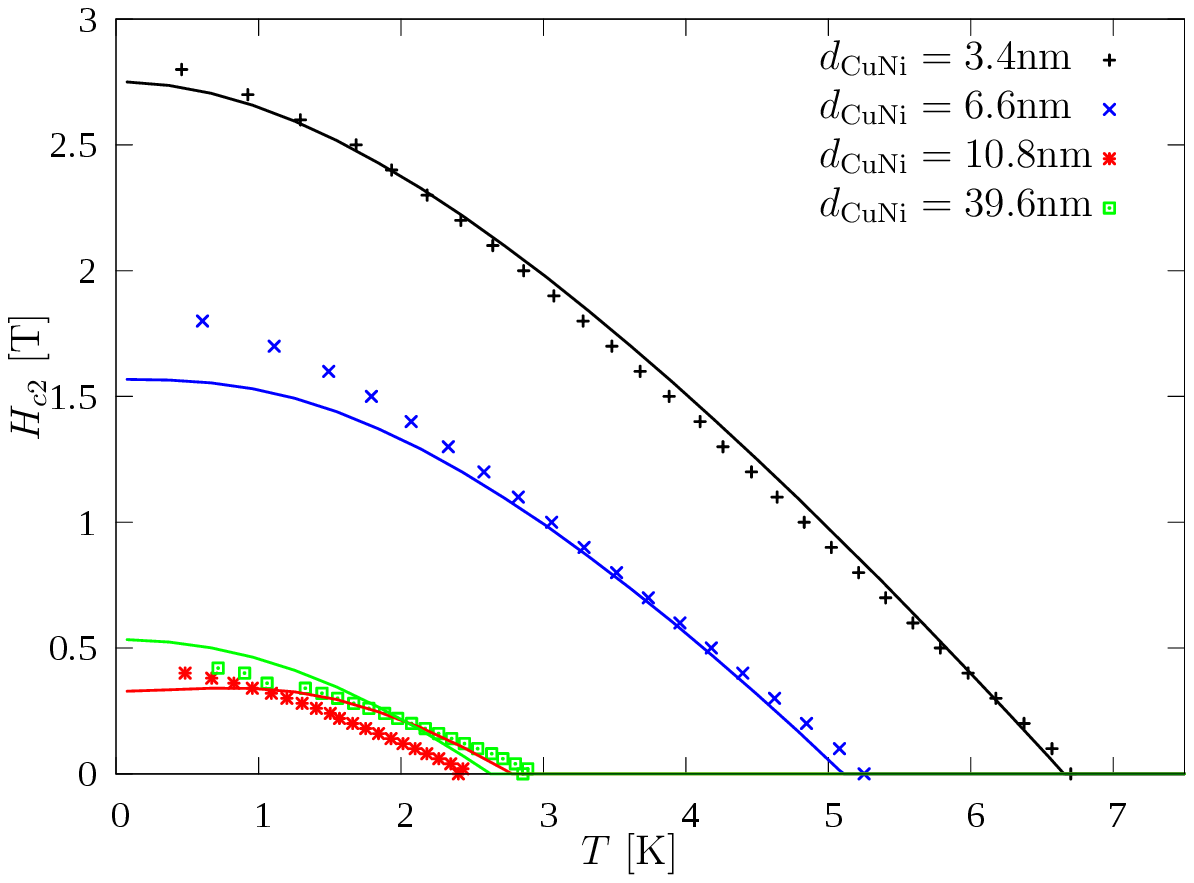}\hfill 
\includegraphics[width=80mm]{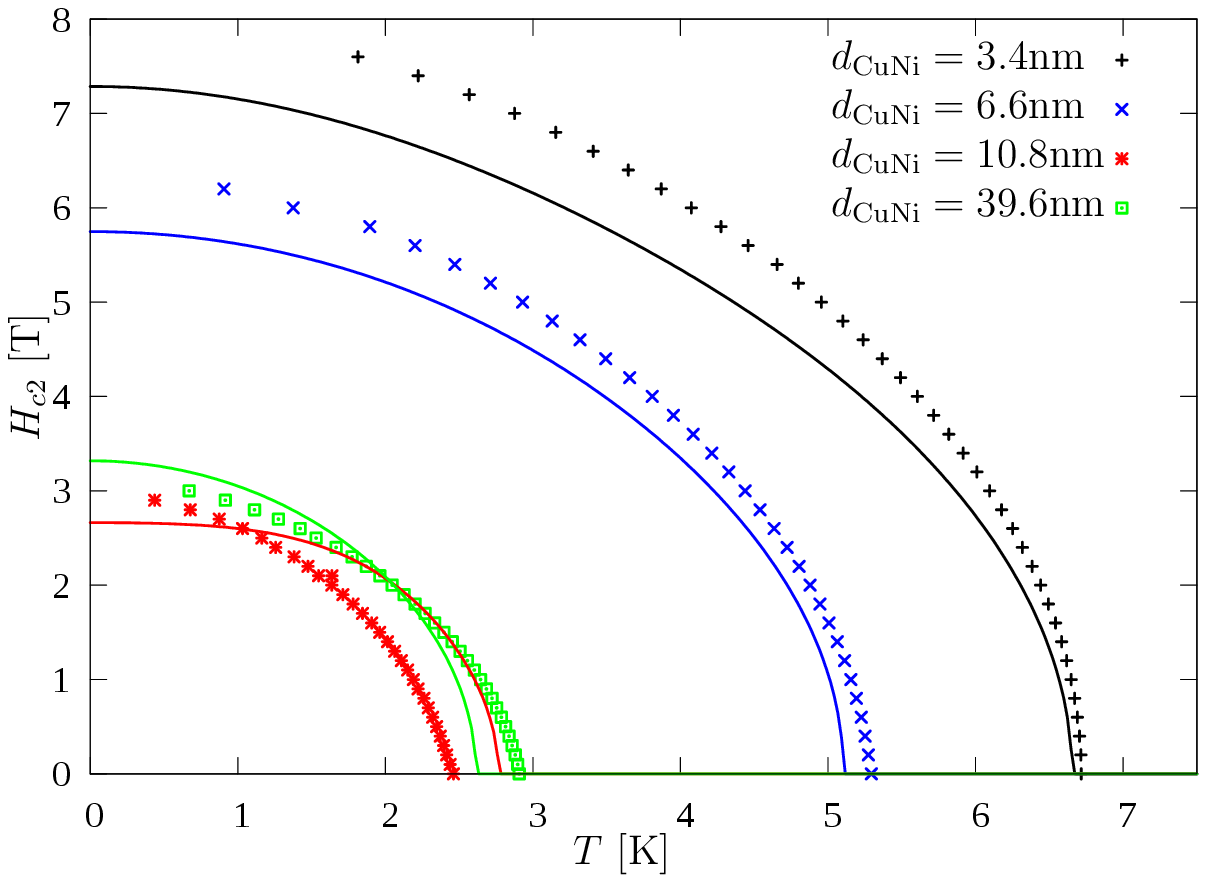}
\caption{Temperature dependence of perpendicular (left panel) and parallel
(right panel) critical magnetic fields for our FSF samples with the niobium
layer thickness value $d_{\mathrm{Nb}}=15.5$\,nm. With a good accuracy the
thicknesses of the ferromagnetic layers are equal, i.e. 
$d_1=d_2=d_{\mathrm{CuNi}}/2$. Experimental data are shown by symbols. 
Solid lines represent fits to our theoretical predictions.}
\label{FSF-HCT-fig}
\end{figure*}

Figure \ref{FSF-HCT-fig} displays the temperature dependencies for
perpendicular (left panel) and parallel (right panel) critical magnetic
fields for the FSF3 series with the niobium layer thickness value
$d_{\mathrm{Nb}}=15.5$~nm. One observes that both critical fields
demonstrate a non-monotonous dependence on the ferromagnetic CuNi 
layer thickness. This observation requires an explanation which will
be obtained by means of our theoretical analysis developed in the next
section.

\section{Theory}

\begin{figure}[tbp]
\centerline{
\includegraphics[width=80mm]{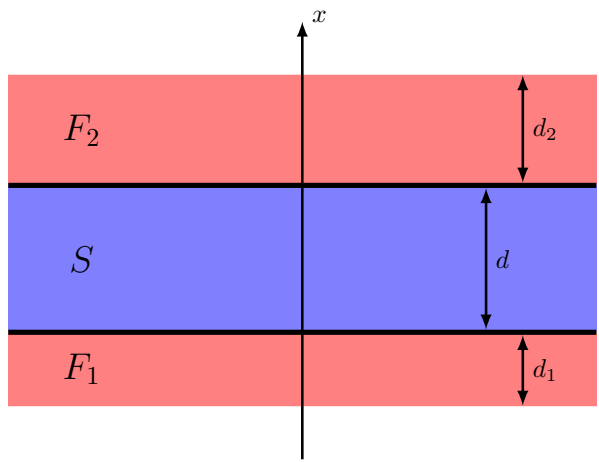}
}
\caption{(Color online) F$_{1}$SF$_{2}$ trilayer structure.}
\label{fsf-fig}
\end{figure}

In this Section we elaborate a theoretical analysis of the superconducting
phase transition in F$_{1}$SF$_{2}$ trilayers in the presence of an external
magnetic field. A sketch of our trilayer is shown in Fig. \ref{fsf-fig}.
Our treatment is based on the quasiclassical theory of superconductivity. As
in the vast majority of experiments the electron elastic mean free path 
$\ell $ in both superconducting and ferromagnetic layers remain much smaller
than the coherence length $\xi _{0}\sim v_{F}/T_{c}$, and $\xi _{F0}\sim v_{F}/h$, 
respectively, where $h$ is the exchange energy, it would be
appropriate to stick to the so-called dirty limit and apply the Usadel
equations \cite{Usadel70,Belzig}. In addition, we will make use of the fact
that at temperatures sufficiently close to the critical one, $T\sim T_{c}$,
the superconducting order parameter $\Delta (T)$ is much smaller than
temperature, $\Delta (T)\ll T$. In this limit, one can linearize the Usadel
equations for the anomalous quasiclassical Green function $F$ and obtain
\begin{equation}
D\left( \nabla -2i\dfrac{e}{c}\bm{A}\right)^{2}F 
-2(|\omega _{n}|+ih\sgn(\omega _{n}))F+2\Delta (\bm{r})=0,  \label{F}
\end{equation}
where $D=v_{F}\ell /3$ is the diffusion constant, $\bm{A}$ is the vector
potential, $\omega _{n}=\pi T(2n+1)$ is the Matsubara frequency and $h$ is
the exchange field in the ferromagnet. The exchange field $h$ and the
superconducting order parameter $\Delta $ differ from zero, respectively, in
the ferromagnetic and superconducting layers only. For simplicity, in what
follows, we will assume that the magnetization in each ferromagnetic layer
is spatially uniform within this layer, and that mutual orientation of the
two magnetizations is collinear, i.e. either parallel (P), or antiparallel
(AP).

Equation (\ref{F}) should be supplemented by appropriate boundary conditions
matching the $F$-functions at the S/F interfaces \cite{Kuprianov88,Garcia2007}. 
In the vicinity of the critical temperature these boundary conditions read
\begin{equation}
r\sigma _{-}\dfrac{\partial F_{-}}{\partial x}=r\sigma _{+}\dfrac{\partial
F_{+}}{\partial x}=F_{+}-F_{-},  \label{bc}
\end{equation}%
where $r$ is the interface resistance per unit area, $\sigma _{\pm }$ and 
$F_{\pm }$ are the Drude conductivities and anomalous Green functions at both
sides of the corresponding S/F interface, respectively. Below we will
consider two orientations of the external magnetic field: perpendicular and
parallel to the S/F interfaces.

\subsection{Perpendicular upper critical field}

Let us first consider an F$_{1}$SF$_{2}$ structure subject to an external
magnetic field $H$ perpendicular to the S/F interfaces. In order to account
for a uniform magnetic field we can choose the gauge $\bm{A}=(0,0,Hy)$,
where $y$, $z$ are in-plane coordinates, and $x$-axis is perpendicular to
the layers' plane. In this case we can use the following Ansatz for both the
anomalous Green function and the order parameter \cite{Radovic91}
\begin{gather}
F(\bm{r},\omega _{n})=g(y,z)F(x,\omega _{n}), \\
\Delta (\bm{r})=g(y,z)\Delta (x),
\end{gather}%
where the function $g(y,z)$ obeys the equation
\begin{equation}
\left( \dfrac{\partial ^{2}}{\partial y^{2}}+\left[ \dfrac{\partial }{
\partial z}-2i\dfrac{e}{c}Hy\right] ^{2}\right) g(y,z)
=-\dfrac{4\pi |H|}{\Phi _{0}}(m+1/2)g(y,z),\quad m\geqslant 0,
\end{equation}
and $\Phi _{0}=\pi c/e$ is the flux quantum. Then, Eq. \eqref{F} takes the
form
\begin{equation}
D\left[ \dfrac{\partial ^{2}}{\partial x^{2}}-\dfrac{2\pi |H|}{\Phi _{0}}
(1+2m)\right] F \\
-2(|\omega _{n}|+ih\sgn(\omega _{n}))F+2\Delta (x)=0.  \label{F_perp}
\end{equation}
Since the magnetic field enters only through the second term in Eq. %
\eqref{F_perp}, in order to obtain the critical field it suffices to
restrict our analysis to the lowest Landau level, \textit{i.e.} to set $m=0$.

In the ferromagnetic regions the superconducting order parameter $\Delta $
equals to zero identically, and Eq. (\ref{F_perp}) can be solved exactly. In
the superconducting region it is necessary to obtain the solution of the
equation
\begin{equation}
D_{S}\left[ \dfrac{\partial ^{2}}{\partial x^{2}}-\dfrac{2\pi |H|}{\Phi _{0}}
\right] F_{S}-2|\omega _{n}|F_{S}+2\Delta (x)=0  \label{F_perp2}
\end{equation}%
with the effective boundary conditions
\begin{gather}
\left. \dfrac{\partial F_{S}}{\partial x}\right\vert _{x=0}=W_{1}^{\perp
}(\omega )F_{S}(x=0),  \label{W1} \\
\left. \dfrac{\partial F_{S}}{\partial x}\right\vert _{x=d}=-W_{2}^{\perp
}(\omega )F_{S}(x=d),  \label{W2}
\end{gather}%
where the parameters $W_{1,2}^{\perp }$ account for the influence of the top
and the bottom ferromagnetic layers, respectively, on the superconducting
layer. They read
\begin{gather}
W_{1,2}^{\perp }=\dfrac{1}{\sigma _{S}}\dfrac{1}{r_{1,2}+\dfrac{\coth
k_{1,2}d_{1,2}}{\sigma _{1,2}k_{1,2}}},  \label{Wdef} \\
k_{1,2}=\sqrt{\dfrac{2}{D_{1,2}}\left( \dfrac{\pi D_{1,2}|H|}{\Phi _{0}}
+|\omega |+ih_{1,2}\sgn(\omega) \right) },  \label{k12def}
\end{gather}%
where $\sigma _{S,1,2}$ and $D_{S,1,2}$ denote the conductivities and the
diffusion constants in the superconductor and in the two ferromagnetic
layers with thicknesses $d$, $d_{1}$, and $d_{2}$, respectively (see Fig. 
\ref{fsf-fig}). Exchange fields of the ferromagnets are denoted by $h_{1}$
and $h_{2}$. The interface resistances per unit area $r_{1}$ and $r_{2}$
refer to the S/F$_{1}$- and S/F$_{2}$-interface resistances.

Spatial dependence of the superconducting order parameter is fixed by the
self-consistency equation
\begin{equation}
\Delta (x)=\lambda \pi T\sum_{|\omega _{n}|<\omega _{c}}F(x,\omega _{n}),
\label{self}
\end{equation}%
where $\lambda >0$ is the BCS coupling constant, and $\omega _{c}$ defines
the high frequency cutoff, which is typically of the order of corresponding
Debye frequency. Introducing the superconducting critical temperature $T_{c0}
$ in the absence of both the external magnetic field and the ferromagnetic
layers one can expel $\lambda $ and $\omega _{c}$ from Eq. \eqref{self} and
get
\begin{equation}
\Delta (x)\ln \dfrac{T_{c0}}{T}=\pi T\sum_{|\omega _{n}|<\omega _{c}}\left[
\dfrac{\Delta (x)}{|\omega _{n}|}-F_{S}(x,\omega _{n})\right] ,
\label{Delta_x}
\end{equation}%
where $T_{c0}=\omega _{c}(2\gamma /\pi )e^{-1/\lambda }$ ($\gamma
=e^{C}\approx 1.781$).

The critical magnetic field $H_{c2}^{\perp }(T)$ is determined by a highest
value of $H$ at which the system of equations \eqref{F_perp2}-\eqref{W2}, 
\eqref{Delta_x} still has a non-trivial solution. In order to proceed
further we will employ the fundamental solution method \cite{Fominov02}. It
is convenient to introduce the function $G(x_{1},x_{2},\omega )$ which
inside the superconductor ($0<x_{1},x_{2}<d$) obeys the following equation
\begin{gather}
\left[ \dfrac{\partial ^{2}}{\partial x_{1}^{2}}-k_{S}^{2}\right]
G(x_{1},x_{2},\omega )+\delta (x_{1}-x_{2})=0,  \label{Geq} \\
k_{S}=\sqrt{\dfrac{2}{D_{S}}\left( \dfrac{\pi D_{S}|H|}{\Phi _{0}}+|\omega
|\right) },  \label{ks2}
\end{gather}%
together with the corresponding boundary conditions at the SF interfaces
\begin{gather}
\left. \dfrac{\partial G(x_{1},x_{2},\omega )}{\partial x_{1}}\right\vert
_{x_{1}=0}=W_{1}^{\perp }G(0,x_{2},\omega ), \\
\left. \dfrac{\partial G(x_{1},x_{2},\omega )}{\partial x_{1}}\right\vert
_{x_{1}=d}=-W_{2}^{\perp }G(d,x_{2},\omega ).  \label{Gbound}
\end{gather}%
Equations \eqref{Geq}-\eqref{Gbound} can be resolved analytically with the result
\begin{multline}
G(x_1,x_2,\omega)=\\=
\begin{cases}
\dfrac{
\left[
\cosh k_S x_1 + W_1^{\perp}(\omega) \dfrac{\sinh k_S x_1}{k_S}
\right]
\left[
\cosh k_S (x_2-d) - W_2^{\perp}(\omega) \dfrac{\sinh k_S (x_2-d)}{k_S}
\right]
}{
\left[
k_S + \dfrac{W_1(\omega) W_2^{\perp}(\omega)}{k_S}
\right]
\sinh k_S d+
\left[ W_1^{\perp}(\omega) + W_2^{\perp}(\omega) \right]\cosh k_S d
}, &x_1 < x_2
\\
\dfrac{
\left[
\cosh k_S (x_1-d) - W_2^{\perp}(\omega) \dfrac{\sinh k_S (x_1-d)}{k_S}
\right]
\left[
\cosh k_S x_2 + W_1^{\perp}(\omega) \dfrac{\sinh k_S x_2}{k_S}
\right]
}{
\left[
k_S + \dfrac{W_1^{\perp}(\omega) W_2^{\perp}(\omega)}{k_S}
\right]
\sinh k_S d+
\left[ W_1^{\perp}(\omega) + W_2^{\perp}(\omega) \right]\cosh k_S d
}, & x_2 < x_1.
\end{cases}
\end{multline}
The solution of Eqs. \eqref{F_perp2}-\eqref{W2} can now be
expressed as a convolution of the function $G(x_{1},x_{2},\omega )$ and the
spatially dependent order parameter $\Delta (x)$:
\begin{equation}
F_{S}(x,\omega )=\dfrac{2}{D_{S}}\int\limits_{0}^{d}G(x,x^{\prime },\omega
)\Delta (x^{\prime })dx^{\prime }.  \label{FS}
\end{equation}%
Combining Eqs. \eqref{Delta_x} and \eqref{FS} we effectively reduce our
problem to the integral equation for the superconducting order parameter
which can be resolved in a straightforward way.

To proceed with numerical solution it is useful to discretize this integral
equation employing the Fourier transformation
\begin{gather}
\Delta (x)=\sum_{m=0}^{\infty }\Delta _{m}f_{m}(x), \\
f_{m}(x)=\cos \dfrac{\pi mx}{d},\quad 0<x<d.
\end{gather}%
Then, the integral equation can be transformed into an infinite system of
linear equations,
\begin{multline}
\Delta _{m_{1}}\int_{0}^{d}f_{m_{1}}^{2}(x)dx\ln \dfrac{T_{c0}}{T} 
=\pi T\sum_{|\omega _{n}|<\omega _{c}}\Biggl[\dfrac{\Delta _{m_{1}}}{|\omega
_{n}|}\int_{0}^{d}f_{m_{1}}^{2}(x)dx-\dfrac{2}{D_{S}}\sum_{m_{2}=0}^{\infty
}\Delta _{m_{2}} \\
\times \int_{0}^{d}f_{m_{1}}(x_{1})G(x_{1},x_{2},\omega
_{n})f_{m_{2}}(x_{2})dx_{1}dx_{2}\Biggr],
\end{multline}%
which can be reduced by a partial summation over the Matsubara frequencies
to a form:
\begin{multline}
\left[ \ln \dfrac{T_{c0}}{T}+\psi \left( \dfrac{1}{2}\right) -\psi \left(
\dfrac{1}{2}+\dfrac{D_{S}|H|}{2\Phi _{0}T}+\dfrac{\pi D_{S}m_{1}^{2}}{4Td^{2}
}\right) \right]  \\
\times \Delta _{m_{1}}(1+\delta _{m_{1},0})=\sum_{m_{2}=0}^{\infty
}g_{m_{1}m_{2}}\Delta _{m_{2}},\quad m_{1}=0\ldots \infty ,  \label{Deltameq}
\end{multline}%
where $\psi (x)$ is digamma function. The matrix elements $g_{m_{1},m_{2}}$
describe the strength of the proximity effect in the superconducting layer.
They are defined by the following expression,
\begin{multline}
g_{m_1 m_2}
=\pi T\sum_{|\omega_n| < \omega_c}
\dfrac{4d^3/D_S}{\left[(k_S d)^2 + (\pi m_1 )^2\right]
\left[(k_S d)^2 + (\pi m_2 )^2\right]}
\\\times
\dfrac{[W_1^{\perp} + (-1)^{m_1 + m_2} W_2^{\perp}]k_S \sinh k_Sd +
W_1^{\perp} W_2^{\perp} [\cosh k_Sd - (-1)^{m_1}][1 + (-1)^{m_1+m_2}]}{
\left[
k_S + \dfrac{W_1^{\perp} W_2^{\perp}}{k_S}
\right]
\sinh k_S d+
\left[ W_1^{\perp} + W_2^{\perp} \right]\cosh k_S d
}.
\end{multline}

The critical magnetic field $H_{c2}^{\perp }(T)$ coincides with the maximal
value of the parameter $H$ provided system of equations \eqref{Deltameq} has
a nontrivial solution. In our numerical calculations an infinite system of
equations \eqref{Deltameq} was truncated to a finite one providing
sufficient accuracy of the solution.

\subsection{Parallel upper critical field}

Let us now turn to a configuration of the external magnetic field $\bm{H}$
oriented parallel to the S/F interfaces. For a uniform magnetic field we
choose the gauge $\bm{A}=(0,Hx,0)$ and employ the following Ansatz for the
anomalous Green function and the order parameter \cite{Yoksan06},
\begin{gather}
F(\bm{r},\omega )=\exp \left( i\dfrac{2\pi Hx_{0}}{\Phi _{0}}y\right)
F(x,\omega ), \\
\Delta (\bm{r})=\exp \left( i\dfrac{2\pi Hx_{0}}{\Phi _{0}}y\right) \Delta
(x),
\end{gather}
where $x_{0}$ is a free parameter which should be chosen in a way to
maximize the critical magnetic field (or temperature). Then, the Usadel
equation \eqref{F} reduces to
\begin{equation}
D\left[ \dfrac{\partial ^{2}}{\partial x^{2}}-\left( \dfrac{2\pi H}{\Phi _{0}
}\right) ^{2}(x-x_{0})^{2}\right] F \\
-2(|\omega _{n}|+ih\sgn(\omega _{n}))F+2\Delta (x)=0.  \label{F_par}
\end{equation}
As before, we can effectively transform the above differential equation into
an integral one:
\begin{equation}
F_{S}(x,\omega )=\dfrac{2}{D_{S}}\int\limits_{0}^{d}G(x,x^{\prime },\omega
)\Delta (x^{\prime })dx^{\prime },  \label{FS2}
\end{equation}
where now
\begin{multline}
G(x_1,x_2,\omega)=
-\dfrac{A_2 B_2}{A_1 B_2 - A_2 B_1}
\dfrac{q_1(x_1) q_1(x_2)}{W}
-\dfrac{A_1 B_1}{A_1 B_2 - A_2 B_1}
\dfrac{q_2(x_1) q_2(x_2)}{W}
\\+
\begin{cases}
\dfrac{A_2 B_1}{A_1 B_2 - A_2 B_1}
\dfrac{q_1(x_1) q_2(x_2)}{W}
+\dfrac{A_1 B_2}{A_1 B_2 - A_2 B_1}
\dfrac{q_2(x_1) q_1(x_2)}{W}, & x_1 < x_2,
\\
\\
\dfrac{A_1 B_2}{A_1 B_2 - A_2 B_1}
\dfrac{q_1(x_1) q_2(x_2)}{W}
+\dfrac{A_2 B_1}{A_1 B_2 - A_2 B_1}
\dfrac{q_2(x_1) q_1(x_2)}{W}, & x_1 > x_2,
\end{cases}
\label{Gpara}
\end{multline}
and $q_{1,2}(x)$ denote two linearly independent solutions of
the equation
\begin{equation}
\left[ \dfrac{\partial ^{2}}{\partial x^{2}}-\left( \dfrac{2\pi H}{\Phi _{0}}
\right) ^{2}(x-x_{0})^{2}-\dfrac{2|\omega _{n}|}{D_{S}}\right] q(x)=0.
\label{qeq}
\end{equation}%
The parameters $A_{1,2}$ and $B_{1,2}$ are defined by the following
relations
\begin{gather}
A_{1}=q_{1}^{\prime }(0)-W_{1}^{\parallel }q_{1}(0),\quad
A_{2}=q_{2}^{\prime }(0)-W_{1}^{\parallel }q_{2}(0), \\
B_{1}=q_{1}^{\prime }(d)+W_{2}^{\parallel }q_{1}(d),\quad
B_{2}=q_{2}^{\prime }(d)+W_{2}^{\parallel }q_{2}(d),
\end{gather}%
and
\begin{equation}
W=q_{1}(x)q_{2}^{\prime }(x)-q_{1}^{\prime }(x)q_{2}(x)
\end{equation}
is the Wronskian for the two solutions $q_{1}(x)$ and $q_{2}(x)$. For
simplicity we will assume that the orbital effects of the magnetic field in
the ferromagnetic layers are small. Then, we can obtain parameters 
$W_{1,2}^{\parallel }$ from Eqs. \eqref{Wdef}, \eqref{k12def} in which the
external magnetic field should be set equal to zero $H\equiv 0$. Combining
Eq. \eqref{Gpara} with the equations derived in Sec.~3A, we derive the upper
critical field $H_{c2}^{\parallel }(T)$ for the parallel orientation of the
magnetic field.

On the basis of the analysis developed above the upper critical field $H_{c2}
$ was evaluated numerically for the both cases of perpendicular 
($H_{c2}^{\perp }(T)$) and parallel ($H_{c2}^{\parallel }(T)$) external
magnetic fields. The corresponding results are presented in the next Section.

\section{Comparison between the theory and experiment}

\begin{figure*}
\includegraphics[width=80mm]{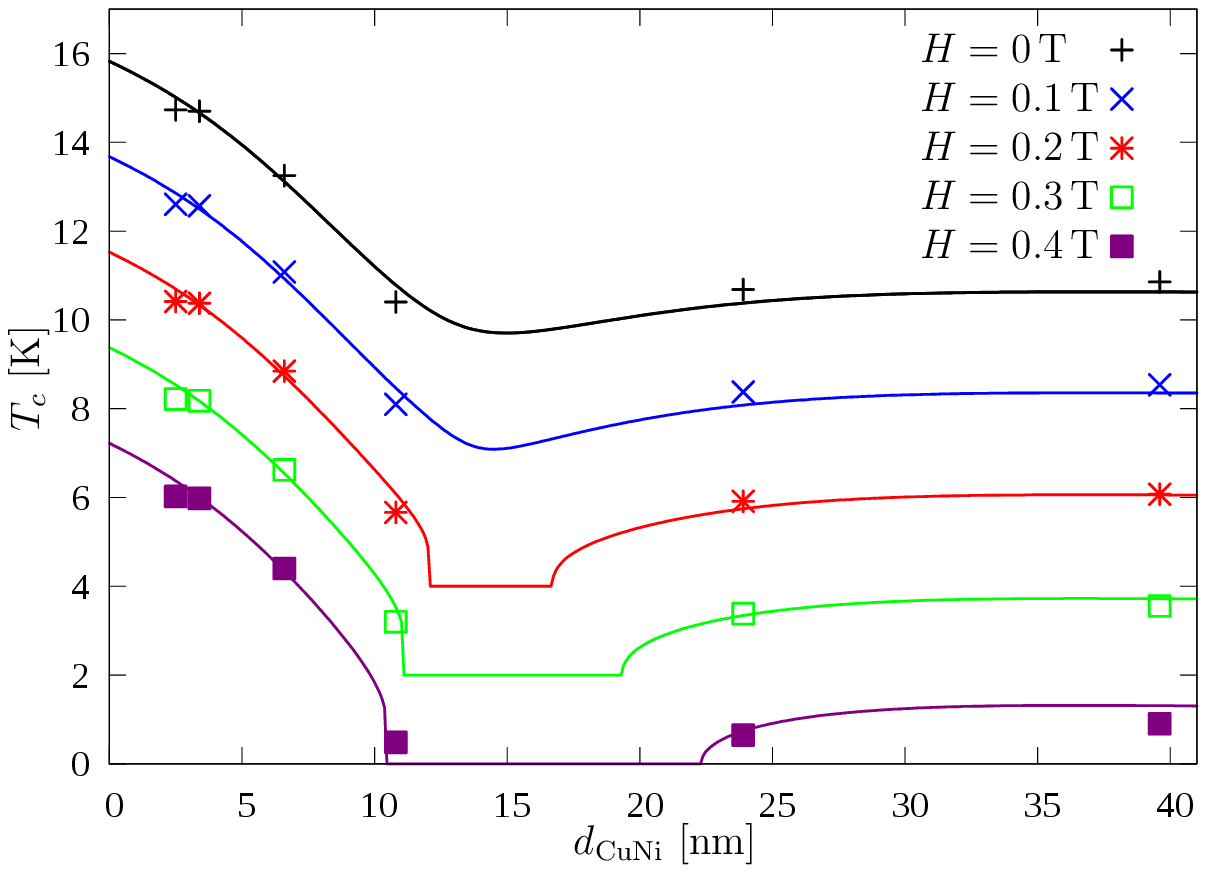}\hfill
\includegraphics[width=80mm]{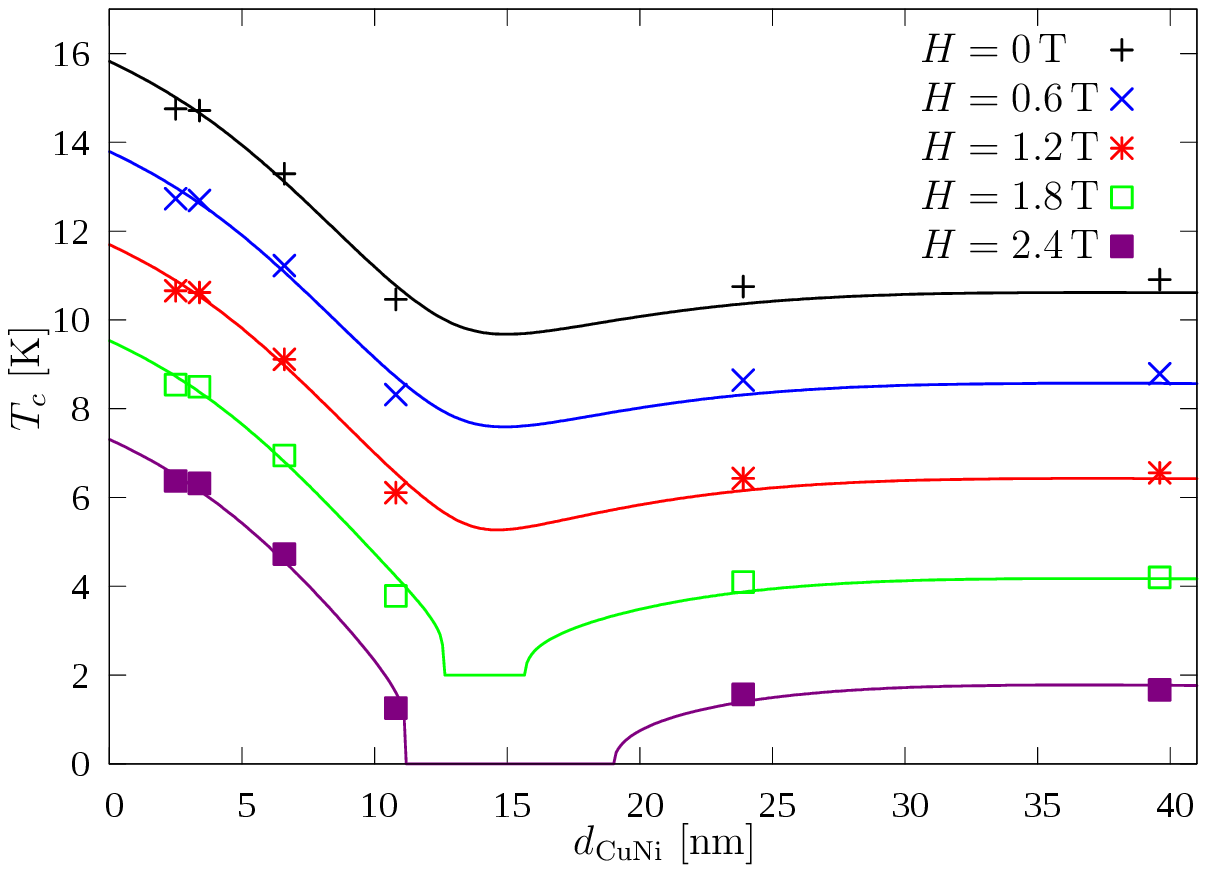}
\caption{Superconducting transition temperature versus the total thickness 
of ferromagnetic layers for both perpendicular (left panel) and parallel 
(right panel) magnetic field orientations. Different curves correspond to different
values of the applied magnetic field (for clarity the curves are shifted upwards 
respectively by  2K, 4K, 6K and 8K). Experimental data points are denoted by symbols.
Solid curves indicate the best fit to our experimental data.}
\label{Tcd-H-fig}
\end{figure*}

Let us perform a detailed comparison between our experimental observations
and theoretical results. By fitting our theoretical curves to the
experimental data it becomes possible to estimate the parameters of our 
F$_{1}$SF$_{2}$ structures which could not be measured directly. Such fits for
the perpendicular critical magnetic field measured in different F$_{1}$SF$_{2}$
 samples are displayed in the left panel of Fig. \ref{FSF-HCT-fig} by
solid lines along with the experimental data points denoted by symbols. The
thickness of the S-layer remains the same for all samples in a series. The
same set of parameters was employed to fit all the experimental data
presented in Fig. \ref{FSF-HCT-fig}. These parameters are:
\begin{gather}
T_{c0}=7.83\,\mathrm{K},\quad \xi _{S}=5.05\,\mathrm{nm},  \label{fit-perp1}
\\
h_{1}=h_{2}=99.4\,\mathrm{K},  \label{fit-perp2} \\
\sigma _{1}=\sigma _{2}=0.38\sigma _{S},\quad D_{1}=D_{2}=3.25D_{S},
\label{fit-perp3}
\end{gather}
Unfortunately, our fitting procedure does not allow to unambiguously
determine the interface resistances $r_{1,2}$. From Eq. \eqref{Wdef} we
observe that under the condition
\begin{equation}
0<r_{1,2}\lesssim \left\vert \dfrac{\coth k_{1,2}d_{1,2}}{\sigma
_{1,2}k_{1,2}}\right\vert
\end{equation}%
the critical temperature remains almost insensitive to the values $r_{1,2}$.
Making use of the fact that the best fit of our data is obtained for fully
transparent interfaces and employing the parameter values \eqref{fit-perp1}-
\eqref{fit-perp3} we arrive at an estimate
\begin{equation}
0\leq r_{1,2}\lesssim \xi _{S}/\sigma _{S}.
\end{equation}

Our theoretical results for the parallel critical magnetic field (solid
lines) are presented in the right panel of Fig. \ref{FSF-HCT-fig} for the
same values of fitting parameters \eqref{fit-perp1}-\eqref{fit-perp3}.

In both cases of perpendicular and parallel magnetic field configurations we
observe a reasonably good agreement between the theory and the experiment,
in particular, having in mind that single set of the fitting parameters was
used for all our samples. Further improvement of the quality of our fits can
be achieved by a small adjustment of these parameters individually for each
sample. We note that one can indeed expect such parameters as exchange field
$h$, conductivity $\sigma $ and diffusion constant $D$ to be slightly
different for the ferromagnetic layers, since these parameters may depend,
\textit{e.g.}, on the layer thickness and different growth conditions for
the bottom and top CuNi layers. There exists also a number of
other physical reasons which might be responsible for a small mismatch
between the theoretical curves and the corresponding experimental data in
Fig. \ref{FSF-HCT-fig}. One of such reasons is that the standard
weak-coupling BCS theory of superconductivity employed here can describe the
properties of niobium samples only approximately \cite{Carbotte90}: strong
coupling corrections to this theory may easily reach $\sim 30\%$ in this
case. This observation alone appears to be sufficient to account for the
remaining discrepancies between the theory and experiment. In addition, the
quasiclassical Usadel equation formalism employed here yields quantitatively
correct results only provided the condition $\ell _{F}\ll \sqrt{D_{F}/h}$ is
fulfilled in ferromagnetic metals. It is not completely clear if this
condition is well satisfied in our samples, which are in the intermediate
regime between strong and weak ferromagnets. Bearing all the above in mind
we conclude that our theory sufficiently well describes the experimental
data for the critical magnetic field in all our samples.

To complete our analysis, in Fig. \ref{Tcd-H-fig} we display both our
theoretical results (solid lines) and the experimental data (symbols) for
the superconducting critical temperature $T_{c}$ as a function of the total
thickness of the ferromagnetic layers at different magnitudes of the applied
magnetic field. A good agreement between the theory and experiment is
observed for all samples. At low magnetic field $T_{c}(d_{\mathrm{CuNi}})$
exhibits nonmonotonous oscillatory dependence which becomes reentrant at high 
enough magnetic fields (the curves from top to bottom in Fig. \ref{Tcd-H-fig},
see also Ref. \cite{Avdeev1}).

In conclusion, we have experimentally and theoretically analyzed the upper
critical field in F$_{1}$SF$_{2}$ trilayers in both perpendicular and parallel
geometries. The series of samples with variable thicknesses of F-layers
were prepared in a single run utilizing the original wedge 
deposition and moving magnetron techniques. The sample series exhibit an 
oscillatory behavior $T_c(d_F)$ in absence of an external magnetic field, 
so several characteristic points in this dependence were chosen 
for the critical fields measurements. 
Perpendicular $H_{c2}^{\perp}(T)$ and parallel $H_{c2}^{\parallel}(T)
$ critical fields were measured in the same samples in the temperature range
$0.4-8$~K and for magnetic fields up to 9 Tesla. A transition 
from the oscillatory to the reentrant behavior of $T_c(d_F)$, 
driven by the external magnetic field, was observed for the first time. 
Temperature dependence
of the critical magnetic field  was derived from
the quasiclassical Usadel equations for arbitrary values
of the system parameters, such as superconducting transition temperature of the
stand-alone superconducting layer $T_{c0}$, superconducting coherence length
$\xi _{S}$, exchange splitting energies $h_{1,2}$ in the F-layers,
conductivities of the ferromagnetic and superconducting layers $\sigma
_{1,2,S}$, conduction diffusion coefficients $D_{1,2,S}$ in the F and S
layers, and the interface resistances $r_{1,2}$ for the S/F$_{1}$- and 
S/F$_{2}$ interfaces, respectively. A fairly good agreement between our theoretical 
predictions and the sets of the experimental data on $H_{c2}^{\perp,\parallel}(T)$ 
for different magnetic layers thickness, as well as $T_c(d_F)$ at different 
magnitudes of the magnetic field, has been achieved for a single set of fit parameters. 
The transition from the oscillatory to the reentrant $T_c(d_F)$ behavior in magnetic 
field has been successfully described by the theory. On one hand, this observation 
demonstrates consistency of the FFLO physics (described within the Usadel 
equation formalism) with the experimental findings obtained in our particular 
Cu$_{41}$Ni$_{59}$/Nb/Cu$_{41}$Ni$_{59}$ trilayer spin-valve cores. 
On the other hand, the theory allows to include more experimental data 
to rectify conditions for obtaining sizable spin-valve effect in 
superconductor-ferromagnet heterostructures.

\section*{Aknowledgements}

We would like to thank V.~V.~Ryazanov, M.~Yu.~Kupriyanov and A.~I.~Buzdin 
for useful discussions as well as M.~Schreck and S.~Gsell for help with RBS measurements.
The work is supported by A.~von~Humboldt Foundation within the frame 
of the Institutspartnerschaften Project and in part by the DFG grant No.~GZ:~HO 955/6-2.

\end{document}